\documentclass[a4paper,12pt]{article}
\usepackage{amsmath}
\usepackage{color} \usepackage{amssymb} \usepackage{graphicx}
\usepackage{axodraw}

\begin{document}
\newcommand{\bsg}{BR(b\rightarrow s \gamma)}
\newcommand{\DM}{\Omega_{CDM}h^2}
\newcommand{\DMW}{\Omega^{WMAP}_{CDM}h^2}
\newcommand{\gmu}{\delta a_{\mu}}
\newcommand{\nn}{\nonumber}
\newcommand{\stau}{\tilde{\tau}}
\newcommand{\sel}{\tilde{e_R}}
\newcommand{\smu}{\tilde{\mu_R}}
\newcommand{\neut}{\tilde{\chi}^0_1}
\newcommand{\nneut}{\tilde{\chi}^0_2}
\newcommand{\nnneut}{\tilde{\chi}^0_3}
\newcommand{\charg}{\tilde{\chi}^+_1}
\newcommand{\ncharg}{\tilde{\chi}^+_2}
\newcommand{\DeltaO}{\Delta^{\Omega}}
\newcommand{\DeltaEW}{\Delta^{EW}}
\newcommand{\dgs}{\delta_{GS}}
\newcommand{\TTBar}{T_2+\overline{T}_2}
\newcommand{\YYBar}{Y_2+\overline{Y}_2}
\newcommand{\sa}{s_{\alpha}}
\newcommand{\mgr}{m_{3/2}}
\newcommand{\tanb}{\tan\beta}
\newcommand{\sigI}{1\sigma}
\newcommand{\sigII}{2\sigma}

\begin{titlepage}

  \begin{center}
    { \sffamily \LARGE Natural Dark Matter from Type I String
      Theory }
    \\[18mm]

    S.~F.~King\footnote{E-mail: \texttt{sfk@hep.phys.soton.ac.uk}},
    and J.~P.~Roberts\footnote{E-mail: \texttt{jpr@phys.soton.ac.uk}}
    \\[10mm]
    {\small\it
      School of Physics and Astronomy,
      University of Southampton,\\
      Southampton, SO17 1BJ, U.K.
    }\\[10mm]
  \end{center}

  \vspace*{0.75cm}

  \begin{abstract}
    We study neutralino dark matter within a semi-realistic type I
    string model, where supersymmetry breaking arises from F-terms of
    moduli fields parameterised in terms of Goldstino angles, which
    automatically gives rise to non-universal soft third sfamily and
    gaugino masses. We study the fine-tuning sensitivities for dark
    matter and electroweak symmetry breaking across the parameter
    space of the type I string model, and compare the results to a
    similar analysis in the non-universal MSSM.  Within the type I
    string model we find that neutralino dark matter can be naturally
    implemented in the $\stau$ bulk region, the $Z^0$ resonance region
    and the maximally tempered Bino/Wino/Higgsino region, in agreement
    with the results of the non-universal MSSM analysis. We also find
    that in the type I string model the ``well-tempered'' Bino/Wino
    region is less fine-tuned than in the MSSM, whereas the $\stau$
    co-annihilation region exhibits a significantly higher degree of
    fine-tuning than in the MSSM.
  \end{abstract}

\end{titlepage}
\newpage
\setcounter{footnote}{0}

\section{Introduction}
\label{Intro}

The prediction of neutralino dark matter is generally regarded as one
of the successes of the Minimal Supersymmetric Standard Model (MSSM).
However the successful regions of parameter space allowed by WMAP and
collider constraints are quite restricted. In a recent paper
\cite{hep-ph/0603095} we discussed fine-tuning with respect to both
dark matter and Electroweak Symmetry Breaking (EWSB) and explored
regions of MSSM parameter space with non-universal gaugino and third
family scalar masses in which neutralino dark matter may be
implemented naturally. For example, we found that the recently
proposed ``well tempered neutralino'' regions \cite{hep-ph/0601041}
involve substantial fine-tuning of MSSM parameters in order to satisfy
the dark matter constraints, although the fine tuning may be
ameliorated if several annihilation channels act simultaneously. To
overcome this we proposed the ``maximally tempered neutralino''
comprising substantial components of Bino, Wino and Higgsino
\cite{hep-ph/0603095}, and showed that it leads to a reduction in
fine-tuning. Moreover we also found other regions of parameter
space which were not ``well tempered'' that exhibit low dark
matter fine tuning. For example the $\stau$ co-annihilation region was
shown to have low fine-tuning, while the bulk region consisting of
t-channel slepton exchange (achievable with non-universal gaugino
masses) was shown to involve no dark matter tuning at all
corresponding to ``supernatural dark matter''.  In all cases
the usual MSSM fine tuning associated with EWSB remained.

Though such a non-universal MSSM provides a general framework for
studying natural dark matter regions, it may not be realistic to
regard the mass terms in the soft supersymmetry (SUSY) breaking
Lagrangian as fundamental inputs since the soft masses merely
parameterise the unknown physics of SUSY breaking. In any realistic
model of SUSY breaking the soft breaking terms in the Lagrangian
should be generated dynamically. It is the parameters that define the
mechanism of SUSY breaking that should be taken as the fundamental
inputs. This immediately raises a difficulty as the true
origin of SUSY breaking is unknown.  In string theory the unknown SUSY
breaking dynamics may be manifested as F-term vacuum expectation
values (VEVs) of hidden sector moduli fields appearing in the theory.
Therefore the values of these F-terms may be regarded as being
more fundamental input parameters than the soft mass terms of the
MSSM. Although the values of the F-terms are unknown, they may be
parameterised in terms of so called Goldstino angles which describe
the relative magnitude of the F-terms associated with the different
moduli fields, as was done for example in type I string theories in
\cite{Ibanez:1998rf}. A more reliable estimate of fine-tuning
sensitivity should therefore result from using such Goldstino
angles, together with the gravitino mass $m_{3/2}$, and some other
undetermined electroweak parameters such as the $\mu$ parameter and
the ratio of Higgs vacuum expectation values $\tan \beta$ as inputs.
Therefore fine-tuning should more properly be calculated with respect
to these inputs. It is possible that fine-tuning when calculated
in terms of such inputs could yield very different results.

In this paper we extend our previous analysis of the non-universal
MSSM to a semi-realistic type I string theory model of the form
originally proposed in \cite{King:2001zi} and phenomenologically
analysed in \cite{hep-ph/0403255} (see also \cite{King:2000ir},
\cite{Benakli:1999jc}, \cite{Shiu:1998pa}, \cite{hep-th/0304200}).
Using such a string model we can address two questions. Firstly, how
does the fine-tuning of a particular dark matter region in the
non-universal MSSM \cite{hep-ph/0603095} compare to a similar region
in the string model? Secondly, do some regions of SUSY breaking
parameter space in the string model more naturally explain dark matter
and electroweak symmetry breaking than others? The model we use to
address these points is the type I string inspired model in
\cite{King:2001zi} in which we can obtain SUSY breaking from any of
twisted (Y) moduli, untwisted (T) moduli or the dilaton (S).  The
phenomenology of SUSY breaking in this model has been studied in
\cite{hep-ph/0403255}. Neutralino dark matter has not so far been
studied in this string model, or any string model involving twisted
moduli, although it has been studied in other string models
\cite{hep-ph/0103218}-\cite{hep-ph/0003186}. However in none of these
cases has the question of the naturalness of the predicted dark matter
density been addressed and, as discussed, one of the main motivations
for the present study is to explore how such results obtained in the
non-universal MSSM translate to the case of a ``more fundamental''
string theory where such non-universality arises automatically.  The
main motivation for revisiting the model in \cite{King:2001zi},
\cite{hep-ph/0403255} is that it exhibits non-universal gaugino masses
and non-universality between the 3rd family and the 1st and 2nd family
squarks and sleptons, which precisely corresponds to the type of
non-universality assumed in \cite{hep-ph/0603095}. This allows a
direct comparison between the non-universal MSSM and a corresponding
type I string model, since the latter shares many of the dark matter
regions previously considered. We will find that dark matter
constraints close off much of the parameter space of the type I string
model, for example the benchmark points suggested in
\cite{hep-ph/0403255} are either ruled out ($\DM\gg
\Omega^{WMAP}_{CDM}h^2$) or disfavoured ($\DM\ll
\Omega^{WMAP}_{CDM}h^2$). However we will find new successful
regions of dark matter in the string model, which mirror some of those
found in the non-universal MSSM, some of which exhibit degrees of
fine-tuning in agreement with the previous results
\cite{hep-ph/0603095}, and some which vary significantly.

The layout of the remainder of this paper is, then, as follows.  In
section \ref{Model} we summarise the string model of
\cite{hep-ph/0403255} and analyse the structures of the GUT scale soft
masses specifically with respect to their implications for dark
matter. In section \ref{Res} we use numerical scans\footnote{As before
  we use {\tt SOFTSUSY v.1.9.1}\cite{hep-ph/0104145} to compute the
  RGE running of the soft parameters and {\tt micrOMEGAs
    v.1.3.6}\cite{hep-ph/0112278} to calculate $\DM$, $\gmu$ and
  $\bsg$} to study the fine-tuning of dark matter within such a model
and find important variations from the general results of
\cite{hep-ph/0603095}. In section \ref{Conc} we present our
conclusions.

\section{The Model}
\label{Model}

\subsection{The brane set-up}

\begin{figure}
  \begin{center}
    \begin{picture}(280,200)(0,0)
      \Line(20,20)(20,180) \Line(20,20)(260,20) \Vertex(20,20){2}
      \Vertex(260,20){2} \LongArrow(140,10)(260,10)
      \LongArrow(120,10)(20,10) \PhotonArc(20,20)(30,0,90){3}{3}
      \PhotonArc(20,20)(40,0,90){3}{3} \PhotonArc(200,110)(20,0,360){3}{6}
      \PhotonArc(150,20)(20,0,180){3}{4} \Text(265,20)[l]{$5_2$}
      \Text(20,185)[b]{$5_1$} \Text(15,15)[tr]{$O$}
      \Text(130,10)[]{$R_{5_2}$} \Text(260,25)[b]{$Y_2$}
      \Text(225,110)[l]{$S,T_i$} \Text(50,50)[bl]{$1,2$}
      \Text(150,45)[b]{$3,H_{u,d}$}
    \end{picture}
    \caption{\footnotesize{The brane set-up from
        \cite{King:2001zi},\cite{hep-ph/0403255}.} \label{branes}}
  \end{center}  
\end{figure}
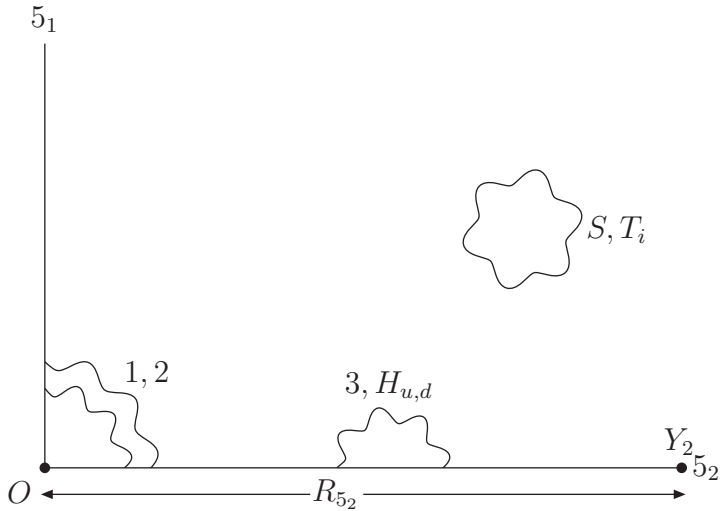

We start with the brane set-up shown in Fig.\ref{branes}, originally
proposed in \cite{King:2001zi},\cite{hep-ph/0403255}. Here we have
two perpendicular intersecting D5 branes $5_1$ and $5_2$. Each holds a
copy of the MSSM gauge group. To maintain gauge coupling unification
at the GUT scale we take the limit of single brane dominance
$R_{5_2}\gg R_{5_1}$. The twisted moduli $Y_2$ is trapped at a fixed
point in the $D5_2$-brane. The untwisted moduli $T_i$ and the dilaton
propagate in the 10D bulk. We identify the first and second family
scalars with open strings with one end on the $5_1$ brane and the
other on the $5_2$ brane. This localises them at the intersection of
the branes and effectively sequesters them from the twisted moduli.
The third family scalars and the Higgs bosons are identified with
strings on the $5_2$ brane.

\begin{figure}
  \begin{center}
    \scalebox{0.8}{\includegraphics{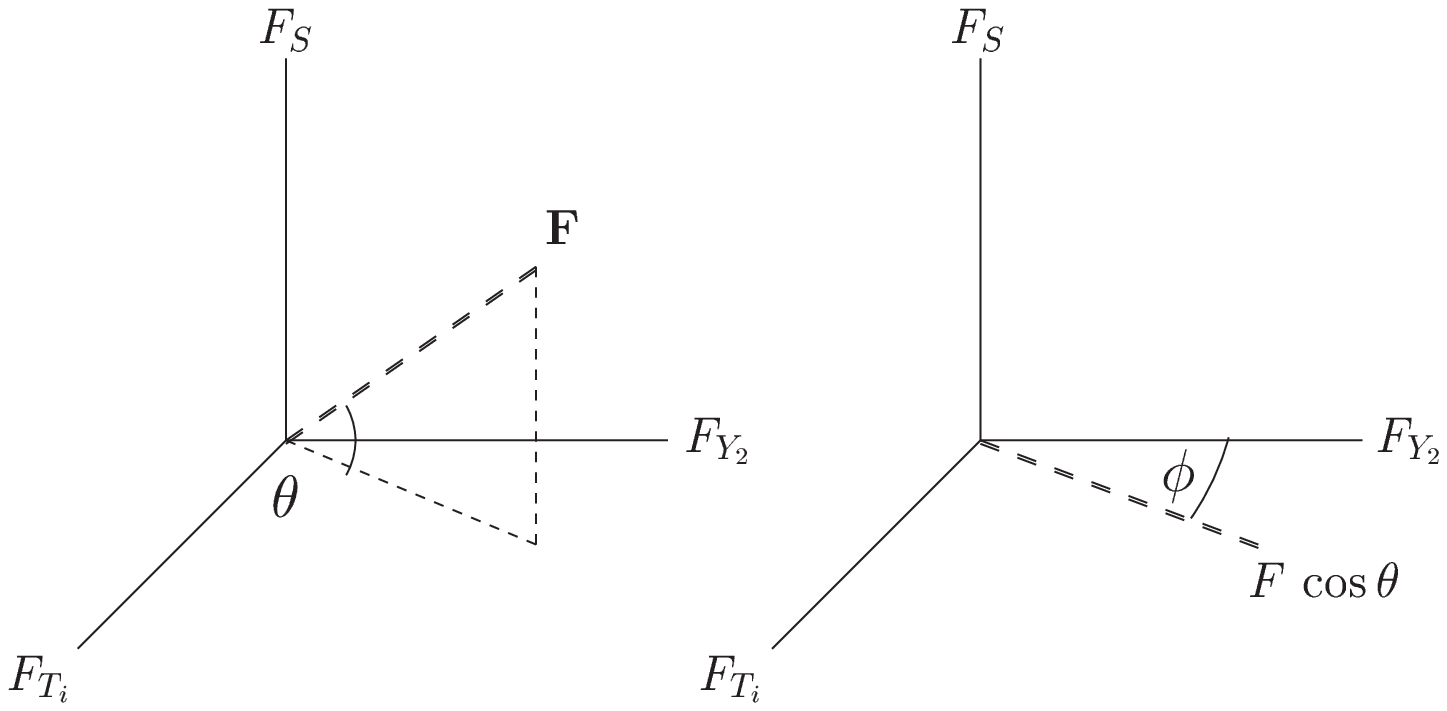}}
  \end{center}
  \caption{\footnotesize{The Goldstino angles are defined to parameterise the
    F-term SUSY breaking coming from the S,T and Y moduli.
    $(\theta,\phi)=(0,0)$ corresponds to twisted moduli ($Y_2$) SUSY
    breaking. $(\theta,\phi)=(0,\pi/2)$ corresponds to untwisted moduli
    ($T_i$) SUSY breaking. $\theta=\pi/2$ corresponds to dilaton ($S$) SUSY
    breaking.}\label{angles}}
\end{figure}

In such a model the SUSY breaking can come from the twisted moduli ($Y_2$)
localised at a fixed point in the $5_2$ brane, the untwisted moduli ($T_i$) in
the bulk or the dilaton ($S$). Each of these forms of SUSY breaking gives
rise to distinct GUT scale soft masses and so to distinct low energy
phenomena. As the exact form of their contribution to the SUSY
breaking F-terms is not known, we use Goldstino angles \cite{Ibanez:1998rf} to
parameterise the relevant contributions of each. These angles are
defined as shown in Fig.\ref{angles}.

\subsection{GUT scale soft masses}

The model determines the soft masses at the GUT scale to be \cite{King:2001zi},\cite{hep-ph/0403255}:
\begin{eqnarray}
  m^2_{\tilde{Q}},m^2_{\tilde{L}},m^2_{\tilde{u}},m^2_{\tilde{d}},m^2_{\tilde{e}}&=&
  \begin{pmatrix}
    m^2_0 & 0 & 0 \\ 0 & m^2_0 & 0 \\ 0 & 0 &
    m^2_{0,3}
  \end{pmatrix} \\
  m^2_{H_u} = m^2_{H_d} &=& m^2_{H}
\end{eqnarray}
where $m^2_0$, $m^2_{0,3}$ and $m_H^2$ are defined:

\begin{eqnarray}
  m_0^2 = m_{3/2}^2 \left[
    1 - \frac{3}{2} \sin^{2} \theta  
    - \frac{1}{2} \cos^{2} \theta \, \sin^{2} \phi  
    - \left( 1-e^{-(T_{2}+\bar{T}_{2})/4} \right) \cos^{2}\theta
    \cos^{2}\phi  \right. \hspace*{15mm} \nonumber\\
  - \frac{X}{3} \, \cos^2 \theta \sin^2 \phi \, \delta_{GS}
  \left( 1- e^{-(T_{2}+\bar{T}_{2})/4} \right) 
  + \frac{X^2}{96} \cos^2 \theta \sin^2 \phi \, 
  e^{-(T_{2}+\bar{T}_{2})/4} (T_{2}+\bar{T}_{2})^{2} \hspace*{5mm}
  \nonumber \\
  - \left. \frac{1}{16 \sqrt{3}} \cos^{2}\theta \cos\phi \sin\phi \, 
    e^{-(T_{2}+\bar{T}_{2})/4} \left\{ 8 (T_{2}+\bar{T}_{2}) +
      \delta_{GS} \, X \right\} X     + {\cal O} \left[ \frac{\delta_{GS} \, 
        e^{-(T_2+\overline{T_2})/4}}{(T_2+\overline{T_2})} \right]\right] \label{m0sq} 
\end{eqnarray}
with $X = Y_2 + \overline{Y_2} - \delta_{GS} \, {\mathrm ln} ( T_2 +
\overline{T_2} )$ where $\dgs$ is the Green-Schwartz parameter.

\begin{eqnarray}
  m^2_{0,3}&=& m_{3/2}^2\left( 1 - \cos^2 \theta \, \sin^2 \phi \right)
  \label{m03sq}\\
  m^2_{H}&=& m_{3/2}^2\left( 1 - 3 \sin^2 \theta \right) 
  \label{mHsq}
\end{eqnarray}

The soft gaugino masses and trilinears are:
\begin{eqnarray}
  M_{\alpha} &=&\frac{\sqrt{3} m_{3/2} \, g_{\alpha}^{2}}{8\pi} \cos \theta
  \left[ \frac{\sin \phi}{\sqrt{3}} 
    \left\{ T_{2} + \bar{T_{2}}
      + \frac{s_{\alpha}}{4\pi} \delta_{GS} \right\}
  \right. \label{Mi} \hspace*{5cm} \\
  &&- \left. \cos \phi \left\{ \frac{\delta_{GS}}{T_{2}+\bar{T}_{2}} -
      \frac{s_{\alpha}}{4\pi} \right\}  + {\cal O} \left[ \left(\frac{\delta_{GS}}{T_2+\overline{T_2}}\right)^2 \right]
  \right] 
  \nonumber\\
  \mathcal{A}  &=& - m_{3/2} \left( \cos\theta \, \sin\phi
   + {\cal O} \left[ \frac{\delta_{GS}}{(T_2 + \overline{T_2})^2} \right]\right)
    \label{eq:softtri}
\end{eqnarray}
where we follow \cite{hep-th/9905098} in taking the parameter
$s_{\alpha}$ to be equal to the MSSM 1-loop $\beta$-function
coefficients: $s_\alpha=\beta_\alpha$ where
$\beta_{\alpha}=2\pi\left\{33/5,1,-3\right\}$. Note that all the soft
masses scale as $\mgr$ as expected in any SUGRA theory.

\subsection{Fine-tuning and the set of input parameters}

The measure we use to study the fine-tuning required to provide
electroweak symmetry breaking is
\cite{Barbieri:1987fn}-\cite{Allanach:2006jc}\footnote{See
  \cite{hep-ph/0603095} for a discussion of the use of these
  sensitivity parameters to measure fine-tuning.}:
\begin{equation}
  \Delta_a^{\text{EW}}=\frac{\partial \ln\left(m_Z^2\right)}{\partial
    \ln\left(a \right)}
  \label{EWmeas}
\end{equation}
Similarly the measure we use to study the fine-tuning of dark matter
is the sensitivity parameter
\cite{hep-ph/0603095},\cite{Ellis:2001zk},\cite{hep-ph/0202110}:
\begin{equation}
  \Delta_a^{\Omega}=\frac{\partial
    \ln\left(\DM\right)}{\partial \ln\left(a \right)}
  \label{meas}
\end{equation}

Clearly the value of $\DeltaO$ depends directly on our choice of
inputs for a theory. In the non-universal MSSM studied previously we
took our inputs at the high energy (GUT) scale as $a=a_{MSSM}$ where:
\begin{equation}
  a_{MSSM}\in \left\{m_0, m_{0,3}, M_1, M_2, M_3, A_0, \tan\beta, \text{sign}(\mu)\right\}
\end{equation}
Here $m_0$ is the soft scalar mass of the first and second family of
squarks and sleptons, $m_{0,3}$ is the soft scalar mass of the third
family of squarks and sleptons and Higgs doublets, $M_i$ are the three
soft gaugino masses, $A_0$ is the universal trilinear soft mass
parameter, $\tan\beta$ is the ratio of Higgs vacuum expectation
values, and $\mu$ is the Higgsino mass parameter.  

Within the present type I string model we take $a=a_{string}$ where:
\begin{equation}
  a_{string}\in\left\{\mgr, \dgs, \TTBar, \YYBar, \theta, \phi, \tanb, \text{sign}(\mu)\right\}
\end{equation}
Here $\tanb$ and sign$(\mu)$ are as in the general MSSM study as they
result from the requirement that the model provide radiative
electroweak symmetry breaking. $\theta$ and $\phi$ are the Goldstino
angles that parameterise the different contributions to SUSY breaking
from the moduli and the dilaton.  The remaining parameters are
directly related to the moduli. The untwisted moduli $T_i$ determine
the radii of compactification.  $\TTBar$ parameterises the
compactification radius in the $5_2$ direction via the
relation\cite{Ibanez:1998rf}:
\begin{equation}
  R_{5_2}=\frac{1}{2}\sqrt{\TTBar}
\end{equation}
As the twisted moduli are trapped at the fixed point at one end of
the brane and the 1st and 2nd families of scalars are trapped at the
other end of the brane, the radius of compactification, and
therefore $\TTBar$, governs the degree of sequestering. This is
evident in the limits of Eq.\ref{m0sq}: as $\TTBar\rightarrow
\infty$, $m^2_0\rightarrow 0$.

Within this paper we follow \cite{hep-ph/0403255} in taking
$\TTBar=50$ and $\YYBar=0$. This maintains the validity of the series
expansion in $\dgs/(\TTBar)$ used to determine the F-terms. However,
as these VeVs are essentially arbitrary, we include them in our set of
parameters for determining dark matter fine-tuning.

$\dgs$ is a model dependent parameter that depends upon the details
of the anomaly cancellation in the twisted sector. This
calculation is beyond the scope of this paper and we set $\dgs=-10$
throughout. However this value can vary and so we include it in our
calculation of fine-tuning parameters.

\subsection{The structure of the neutralino}

The principle factors in the determination of the dark matter relic
density are the mass and composition of the lightest neutralino. This
is determined by the ratio between $M_1$, $M_2$ and $\mu$ at the low
energy scale. Though we cannot predict the size of $\mu$ from the form
of the soft masses, we can find $M_1$ and $M_2$. The values of $M_i$
at $m_{GUT}$ can be simplified from Eq.\ref{Mi} once we have set
$\TTBar$ and $\dgs$:

\begin{eqnarray}
\nonumber M_1 &=& 0.03 \mgr \cos \theta \left(5.7\sin\phi+3.5\cos\phi\right)\\
M_2 &=& 0.03 \mgr \cos \theta \left(26\sin\phi+0.7\cos\phi\right) \label{simpMi}\\
\nonumber M_3 &=& 0.03 \mgr \cos \theta \left(38\sin\phi-1.3\cos\phi\right)
\end{eqnarray}

\begin{table}
  \begin{center}
    \begin{tabular}{|l|l|c|}
      \hline
      Region & $\phi$ &  $M_1:M_2:M_3$\\
      \hline
      Twisted moduli ($Y_2$) dominated & 0 & $3.5:0.7:-1.3$\\
      Untwisted moduli ($T_i$) dominated & $\pi/2$ & $5.7:26:38$\\
      \hline
    \end{tabular}
  \end{center}
\caption{\footnotesize{The ratio of the GUT scale gaugino masses in the twisted
    moduli ($Y_2$) and untwisted moduli ($T_i$) SUSY breaking
    limits.} \label{MiHier}}
\end{table}

The overall magnitude of the gaugino masses is set by $\mgr$ and
$\cos\theta$. The ratio of GUT scale gaugino masses is determined by
$\phi$, as shown in Table \ref{MiHier}. To analyse the low energy
gaugino mass ratio, and so study the composition of the $\neut$, we
can use the rule of thumb\footnote{The exact relation between the GUT
  scale and low energy masses is determined by the RGEs. We can use
  this simple rule of thumb in the case of the gauginos because their
  one-loop RGEs are straightforward, for their explicit form see
  \cite{hep-ph/0312378}} that $M_1(M_{SUSY})\approx 0.4M_1(m_{GUT})$
and $M_2(M_{SUSY})\approx 0.8M_2(m_{GUT})$. This allows us to see that
in the twisted moduli dominated limit, in the absence of small $\mu$,
we have Wino dark matter. In the untwisted moduli dominated limit,
again without small $\mu$, we have Bino dark matter.  To find the
Wino/Bino well-tempered region we need to find the value of $\phi$
that gives $M_1(m_{SUSY})\approx M_2(m_{SUSY})$. This occurs when
$M_1(m_{GUT})\approx 2M_2(m_{GUT})$ and so the switch from Bino to
Wino dark matter will occur around $\phi\approx 0.05$. Therefore to
study Wino/Bino ``well-tempered'' dark matter we should consider low
values of $\phi$.  At lower values of $\phi$ dark matter will be Wino
and so will annihilate too efficiently to explain the observed dark
matter. At larger $\phi$, dark matter will be Bino or Bino/Higgsino.

In Table \ref{MiHier} we have not included the dilaton dominated limit
$\theta=\pi/2$ for two reasons. Firstly, as $\theta\rightarrow\pi/2$,
$M_i\rightarrow 0$ and the parameter space will be ruled out by LEP
bounds on the neutralinos, charginos and the gluino. As $\cos\theta$
is a common coefficient, the degree of dilaton contribution only
affects the overall mass scale of the gauginos, not their composition.
Secondly we are forbidden from accessing $\theta=\pi/2$, the dilaton
dominated limit, by Eq.\ref{mHsq}. Within this paper we keep the
squared Higgs mass positive at the GUT scale and so limit our studies
to $\theta<sin^{-1}\left(1/\sqrt{3}\right)$. Therefore the dilaton
contribution can only suppress the gaugino masses by a factor of $0.8$
at the most. The primary effect of $\theta$ on the phenomenology is
through the sfermion and Higgs masses.

\begin{table}
  \begin{center}
    \begin{tabular}{|l|l|l|l|l|l|l|}
      \hline
      Point & $\theta$ & $\phi$ & $\mgr$(TeV) & $\tanb$ & $\neut$ & $\DM$\\
      \hline
      A & 0   & 0   & 5 & 4  & Wino & $\DM \ll \DMW$\\
      B & 0.1 & 0.1 & 2 & 10 & Bino & $\DM \gg \DMW$\\
      C & 0.6 & 0.1 & 2 & 20 & Bino & $\DM \gg \DMW$\\
      \hline
    \end{tabular}
  \end{center}
  \caption{\footnotesize{Benchmark points from \cite{hep-ph/0403255}. B and C
    overclose the universe and so are ruled out by dark matter. A lies
    in a region inaccessible within our studies as the parameter space
    has disappeared for $m_t=172.7\text{ GeV}$. However even if the
    parameter space were allowed, the LSP would be Wino and so could
    not reproduce the observed dark matter density.}\label{Bench}}
\end{table}

By considering the structures of the neutralino masses we can quickly
analyse the implications of dark matter for the benchmark points
proposed in \cite{hep-ph/0403255}. In Table \ref{Bench} we list the
soft parameters that define the three benchmark points and note the
resulting composition of the LSP. Point A corresponds to the twisted
moduli dominated limit and the LSP is Wino. Wino dark matter
annihilates efficiently in the early universe resulting in a relic
density far lower than that observed today. For point A to remain
valid, there would have to be non-thermal production of SUSY dark
matter or some other, non-SUSY, particle responsible for the observed
relic density\footnote{As we will show in section \ref{Y2res}, this
  point is also ruled out by LEP bounds on the lightest Higgs if we
  take $m_t=172.7\text{ GeV}$, as we do throughout this paper.}.

Points B and C both result in Bino dark matter. In general Bino dark
matter does not annihilate efficiently, often resulting in a relic
density much greater than that observed. For the density to be in
agreement with the measured density, certain annihilation channels
need to be enhanced. This can happen if (i) the NLSP is close in mass to
the neutralino, allowing for coannihilation, (ii) neutralinos can
annihilate to a real on-shell Higgs or $Z$ or (iii) there exist light
sfermions that can mediate neutralino annihilation via t-channel
sfermion exchange. None of these mechanisms exist in the case of
points B or C, resulting in a predicted dark matter density far in
excess of that measured by WMAP.

As the previously proposed benchmark points fail, we go on to scan the
parameter space to find points that agree with the WMAP measurement
of $\DM$.

\section{Results}
\label{Res}

\subsection{Twisted moduli dominated SUSY breaking}
\label{Y2res}

\begin{table}
  \begin{center}
    \begin{tabular}{|l|r|}
      \hline
      Soft Mass & Value \\
      \hline
      $m_0$ & $3.7\times 10^{-6}~\mgr$ \\
      $m_{0,3}$ & $\mgr$ \\
      $m_H$ & $\mgr$ \\
      $M_1$ & $0.1~\mgr$ \\
      $M_2$ & $0.02~\mgr$ \\
      $M_3$ & $-0.04~\mgr$ \\
      $A$ & $0$ \\
      \hline
    \end{tabular}
  \end{center}
  \caption{\footnotesize{In the twisted moduli ($Y_2$) dominated limit,
      $\theta=\phi=0$, the soft masses take the form shown. This limit
      is characterised by the exponential suppression of the 1st and
      2nd family scalar soft masses and a light Wino LSP.}\label{Y2tab}}
\end{table}

In the twisted moduli dominated limit ($\theta=\phi=0$) the soft
masses simplify to the values shown in Table \ref{Y2tab}. In this
regime the 1st and 2nd family scalars have exponentially suppressed
soft masses due to their sequestering from the twisted moduli. The
third family scalars and the Higgs bosons have a universal soft mass
equal to $\mgr$. Finally the lightest neutralino is Wino and very
light.

\begin{figure}[t]
  \begin{minipage}{0.49\textwidth}
    \scalebox{.7}{\includegraphics{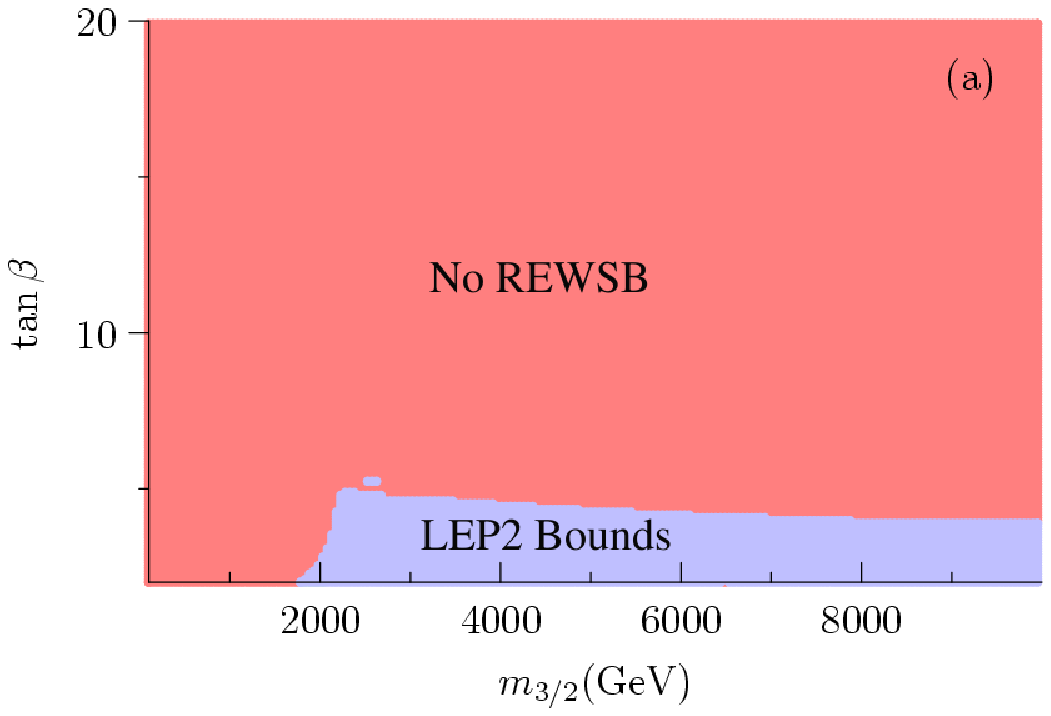}}
  \end{minipage}
  \hspace*{0.5cm}
  \hfill
  \begin{minipage}{0.49\textwidth}
    \scalebox{.7}{\includegraphics{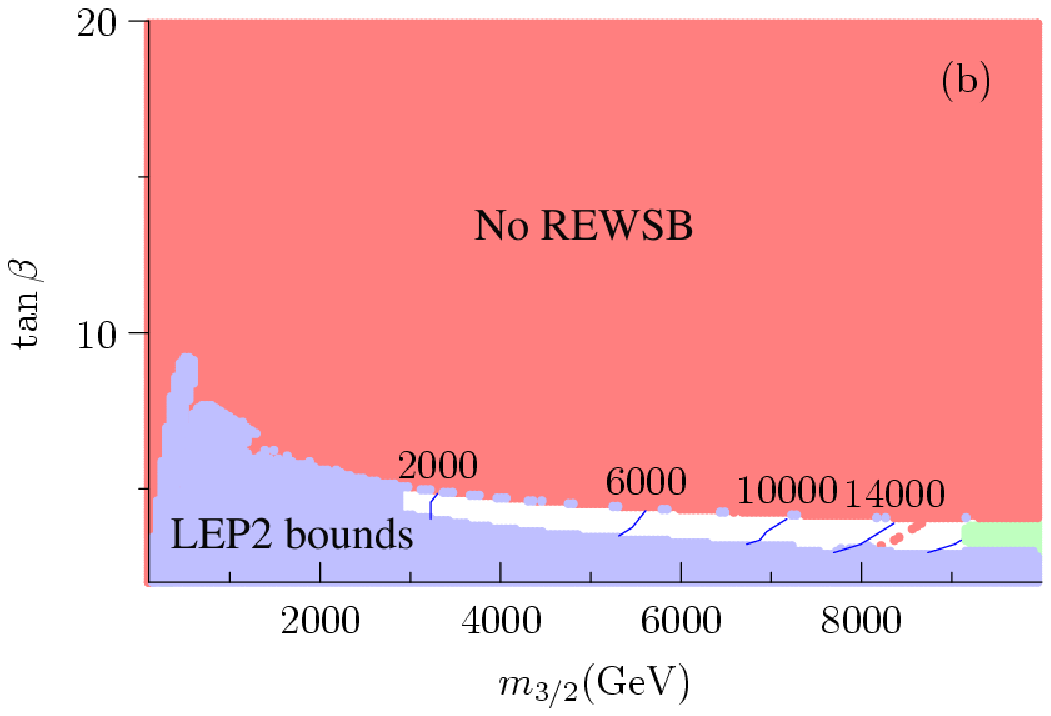}}
  \end{minipage}

  \vspace*{0.5cm}
  \begin{minipage}{0.49\textwidth}
    \scalebox{.7}{\includegraphics{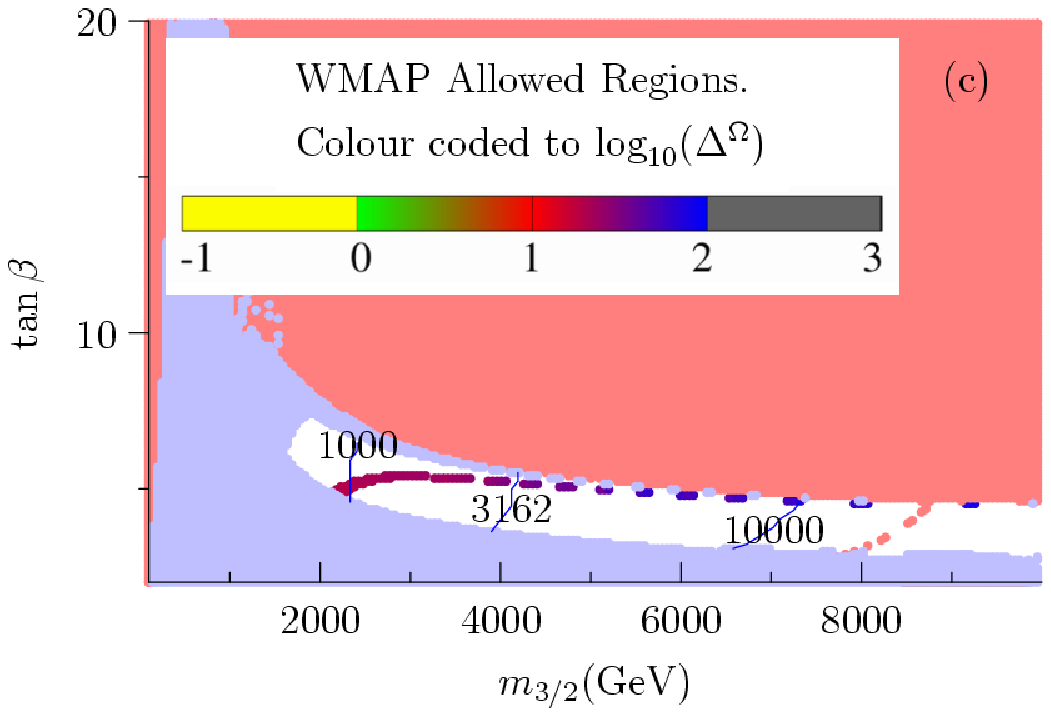}}
  \end{minipage}
  \hspace*{0.5cm}
  \hfill
  \begin{minipage}{0.49\textwidth}
    \scalebox{0.7}{\includegraphics{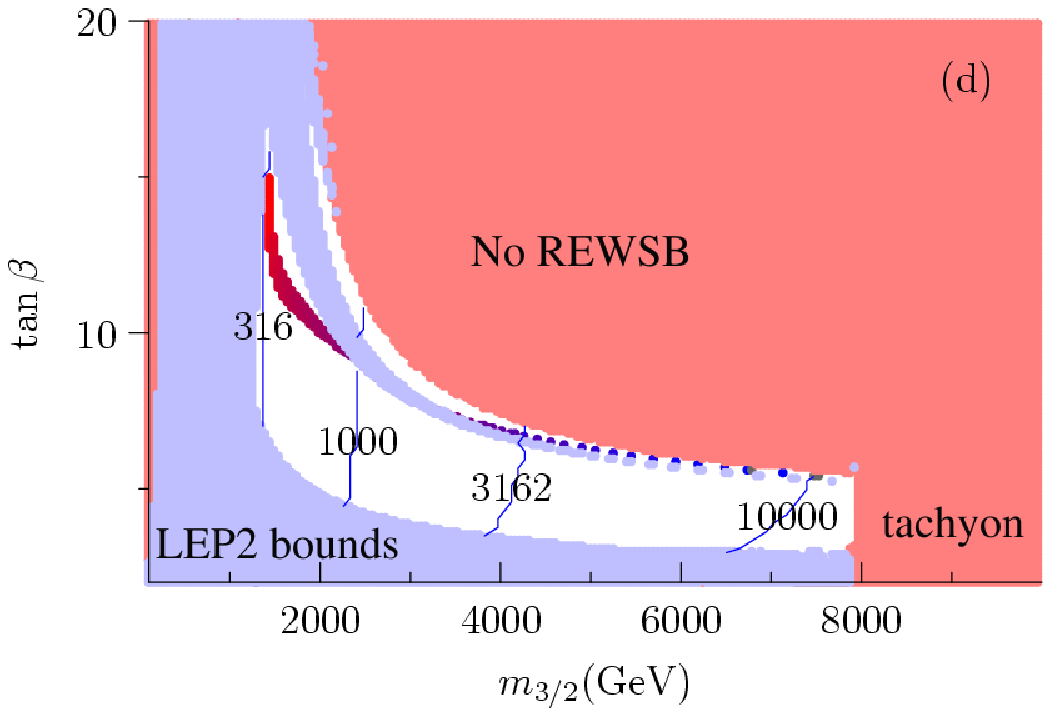}}
  \end{minipage}

  \caption{\footnotesize{Panel (a) shows the twisted moduli dominated limit
      $\theta=\phi=0$. As we switch on contributions from T-moduli,
      the LEP and REWSB bounds recede. In (b) $\phi=0.05$, (c)
      $\phi=0.07$ and (d) $\phi=0.1$. $\theta=0$ throughout. In panels
      (c) and (d) there are regions allowed by WMAP.  These regions
      are plotted in varying colours corresponding to the degree of
      fine-tuning they require. In panel (c) we present a legend for
      this colour coding. Finally, we represent EW tuning by contours
      in panels (b)-(d). $\bsg$ agrees with measurement at $\sigI$
      across the open parameter space but $(g-2)_\mu$ agrees with the
      Standard Model value. The low energy SUSY spectra corresponding
      to these panels are discussed in \cite{hep-ph/0403255}}
    \label{Y2figs}}
\end{figure}

In Figs.\ref{Y2figs}(a)-(d) we examine the phenomenology of the
parameter space as T-moduli contributions are gradually switched on by
slowly increasing $\phi$ from $0$. In the twisted moduli dominated
limit (Fig.\ref{Y2figs}(a)) the parameter space is either closed off
by LEP bounds on the lightest Higgs and chargino or because $\mu^2<0$,
resulting in a failure of radiative electroweak symmetry breaking.
This disagrees with \cite{hep-ph/0403255} because we take
$m_t=172.7\text{ GeV}$ as opposed to $m_t=178\text{ GeV}$.  Therefore
the twisted moduli dominated limit is ruled out by experimental bounds
for the present top mass. 

In Fig.\ref{Y2figs}(b)-(d) we take incrementally larger values of
$\phi=0.05$, $0.07$ and $0.1$ respectively. This has three primary
effects. Firstly $M_2$ increases, and to a lesser extent so does $M_1$
from Eq.\ref{simpMi}. This changes the LSP from Wino to Bino and
quickly increases the mass of the charginos, helping to satisfy LEP
bounds. Secondly the 1st and 2nd family soft scalar masses receive a
substantial contribution from the T-moduli from Eq.\ref{m0sq}. Finally
$M_3$ becomes positive and then steadily increases in size, helping to
mitigate the bounds from REWSB and from the LEP bounds on the lightest
Higgs boson.

The combination of these effects opens up the parameter space as we
increase $\phi$, where the area of parameter space consistent with
collider phenomenology is shown as white space in the figures, and
within this white space the area consistent with WMAP allowed
neutralino dark matter is shown as thin coloured bands, where the
colour coding corresponds to the degree of fine-tuning as explained in
the figure caption.  The first evidence of the model providing a dark
matter density in agreement with that measured by WMAP is in
Figs.\ref{Y2figs}(c) and \ref{Y2figs}(d). In both of these scans, if
$\mu$ were large the LSP would be Bino, with a small proportion of
Wino. However as much of the parameter space is closed off because
$\mu^2<0$, along the edge of this region $\mu$ will be of a comparable
magnitude to $M_1$ resulting in ``well-tempered'' Bino/Higgsino dark
matter. In such regions, co-annihilation with $\nneut$ and $\charg$
become significant and reduces the dark matter density to the
magnitude observed. However the well-tempered region visible at
$4-8$\text{TeV} is plotted in dark blue, corresponding to a
fine-tuning $\DeltaO\approx60$. This is comparable in magnitude to
that of the focus point of the CMSSM. As $\mu$ is sensitive to
$\tan\beta$ and $M_1$ is not, there is no reason for these masses to
be correlated as is required for Higgsino/Bino dark matter.  Therefore
it is unsurprising that the tuning is large and the majority of the
tuning is due to $\tan\beta$, which strongly affects the
calculation of $\mu$.

As we move to lower values of $\mgr$, the colour of the dark matter
strip moves from blue to red. This corresponds to a drop in $\DeltaO$.
To understand this we need to once again consider the composition of
the LSP. Away from the region with low $\mu$, the neutralino is
primarily Bino with a small but significant Wino component. This
results in $\nneut$ and $\charg$ being slightly heavier than $\neut$.
Across much of the parameter space this mass difference is large
enough that co-annihilation effects are unimportant. However, as the
overall mass scale drops, so does the absolute value of the mass
difference between the LSP and the NLSPs.  Below $\mgr=4\text{ TeV}$,
the mass difference is small enough for there to be an appreciable
number density of $\charg$ and $\nneut$ at freeze out to co-annihilate
with the LSP. The efficiency of coannihilation is primarily sensitive
to the mass difference between the LSP and the NLSP. This mass
difference scales slowly with $\mgr$ resulting in a Wino/Bino
well-tempered region that exhibits low fine-tuning $\DeltaO\approx
10$, lower than the tuning required for Wino/Bino regions in
\cite{hep-ph/0603095}.

In Fig.\ref{Y2figs}(b), though there is a region of parameter space
that satisfies LEP bounds and REWSB, there is no WMAP allowed strip.
This is because here the Wino component of the LSP is already too
large and dark matter annihilates too efficiently in the early
universe. This is unfortunate as it is only for low $\phi$ that we
have exponentially suppressed soft masses for the 1st and 2nd
families. We would like to be able to access such a region of
parameter space as light 1st and 2nd family sleptons can provide
neutralino annihilation via t-channel slepton exchange. In
\cite{hep-ph/0603095} we found these regions exhibited very low
fine-tuning. Such a region is not available in this string model
because as soon as we move away from $\phi=0$ the first and second
families gain substantial masses. As soon as we can access Bino dark
matter, the sleptons are already too heavy to contribute significantly
to neutralino annihilation. Though we fail to find a light slepton
bulk region in this limit, in the limit of untwisted moduli dominated
SUSY breaking we will find a light $\stau$ bulk region.

Finally we note that the electroweak fine-tuning is large right across
this parameter space. This is a direct result of the large values of
$\mgr$ that are required to satisfy LEP bounds. When $\phi=0$,
$M_2=0.02\mgr$ from Eq.\ref{simpMi} and charginos are too light. As we
increase $\phi$, the coefficient of proportionality between $M_2$ and
$\mgr$ increases but remains small for small $\phi$. To reach low
$\mgr$ we need to move to regimes in which
$\sin\phi\approx\mathcal{O}(1)$, away from the twisted moduli
dominated limit. These large values of $\mgr$ are responsible for
large electroweak tuning. As $m^2_{0,3}\approx\mgr^2$, the masses
going into our calculation of electroweak symmetry breaking are
$\mathcal{O}(\mgr)$. We need to tune our soft masses to cancel to
provide the correct value of $m_Z$, orders of magnitude lighter. As we
increase $\mgr$ we increase the degree of fine-tuning required. To
access regions with low fine-tuning we need to access low $\mgr$, and
that means taking large $\phi$, as we consider next.

\subsection{T-moduli dominated SUSY breaking}

\begin{table}
  \begin{center}
    \begin{tabular}{|l|r|}
      \hline
      Soft Mass & Value \\
      \hline
      $m_0$ & $131~\mgr$ \\
      $m_{0,3}$ & $0$ \\
      $m_H$ & $\mgr$ \\
      $M_1$ & $0.17~\mgr$ \\
      $M_2$ & $0.78~\mgr$ \\
      $M_3$ & $1.14~\mgr$ \\
      $A$ & $-\mgr$ \\
      \hline
    \end{tabular}
  \end{center}
  \caption{\footnotesize{The soft masses in the untwisted moduli ($T_i$) dominated
      limit, $\theta=0, \phi=\pi/2$. This limit is characterised by
      vanishing 3rd family scalar masses and a Bino LSP.}\label{T2tab}}
\end{table}

In the limit in which all the SUSY breaking comes from the untwisted
T-moduli $(\theta=0$, $\phi=\pi/2)$, the soft masses take the form shown
in Table \ref{T2tab}. In the gaugino sector, as $M_1<M_2$, the
lightest neutralino will have no Wino component. Unless there is a
part of the parameter space with low $\mu$, the LSP will be Bino. As
Bino dark matter on its own generally annihilates extremely
inefficiently there would need to be other contributions to the
annihilation cross-section to satisfy WMAP bounds. The other defining
feature of this limit is that $m_{0,3}=0$. As the third family
particles all pick up masses through loop corrections, they will not
be massless at the low energy scale.  However these corrections are
smallest for $\stau_1$ and will leave it light. This
opens up the possibility that t-channel stau exchange and stau
co-annihilation will help to suppress the Bino dark matter density.

As the 1st and 2nd family particles have a large soft mass, they will
not provide a contribution to the muon $(g-2)$ value. Therefore this limit
will not agree with the measured deviation $\gmu$ from the standard
model value\cite{hep-ph/0409360}. In this limit, the model predicts a
value of $(g-2)_{\mu}$ in agreement with the Standard Model.

\begin{figure}[t]
  \begin{minipage}{0.49\textwidth}
    \scalebox{.7}{\includegraphics{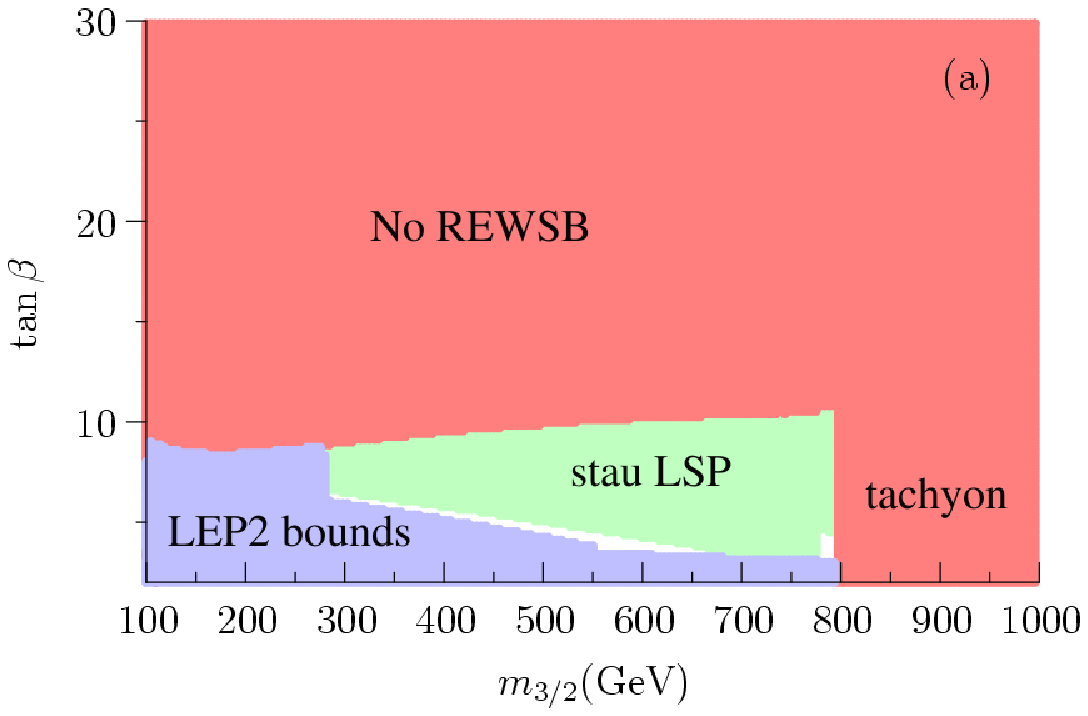}}
  \end{minipage}
  \hspace*{0.5cm}
  \hfill
  \begin{minipage}{0.49\textwidth}
    \scalebox{.7}{\includegraphics{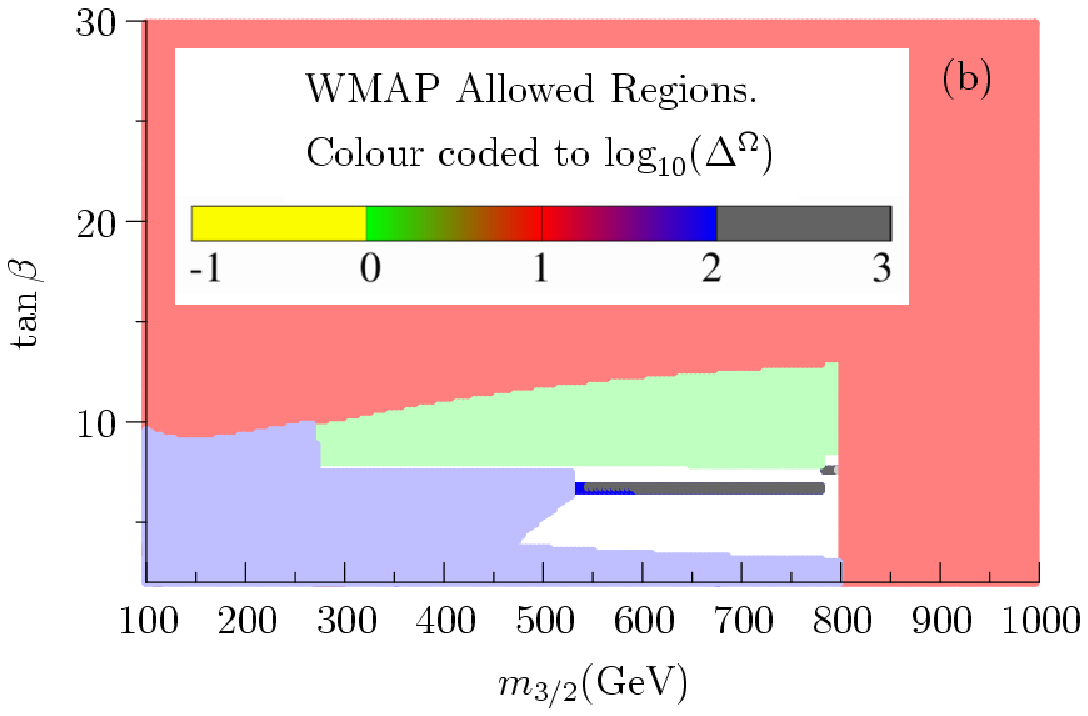}}
  \end{minipage}

  \vspace*{0.5cm}
  \begin{minipage}{0.49\textwidth}
    \scalebox{.7}{\includegraphics{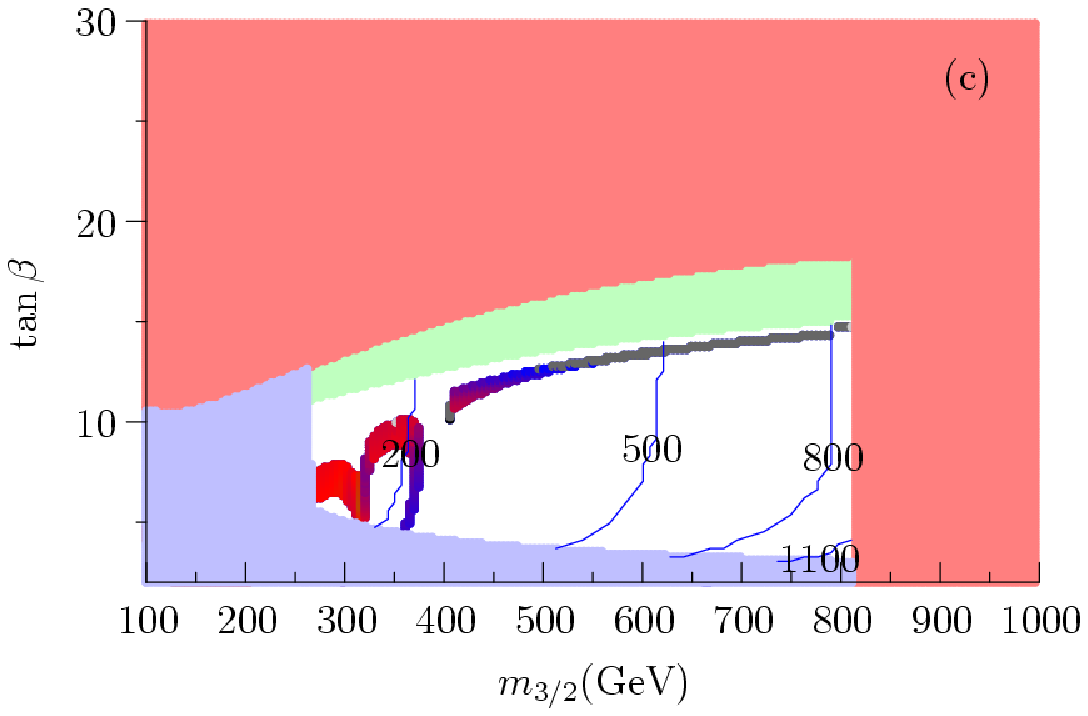}}
  \end{minipage}
  \hspace*{0.5cm}
  \hfill
  \begin{minipage}{0.49\textwidth}
    \scalebox{0.7}{\includegraphics{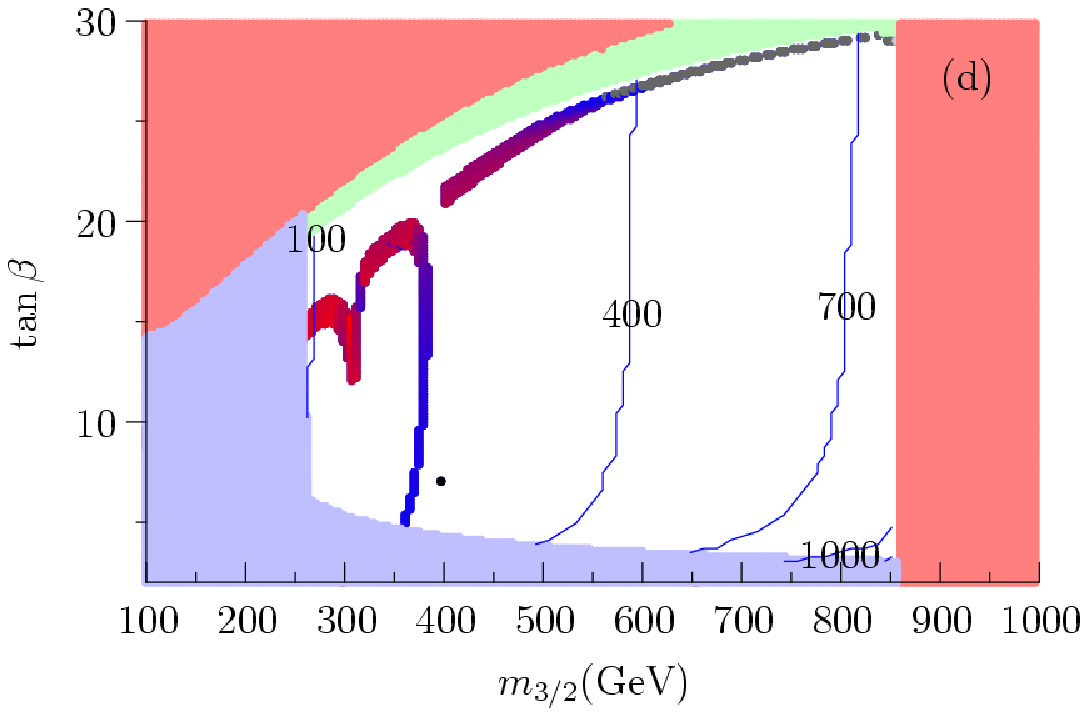}}
  \end{minipage}

  \caption{\footnotesize{Panel (a) shows the T-moduli dominated limit 
      $\theta=0,\phi=\pi/2$ in which the parameter space is entirely
      closed off by experimental bounds. As soon as we move away from
      $\phi=\pi/2$, the parameter space opens up and we find dark
      matter allowed regions. In (b) $\phi=15\pi/32$, (c)
      $\phi=7\pi/16$ and (d) $\phi=3\pi/8$.  Once again we switch off
      the dilaton contributions by taking $\theta=0$ throughout. In
      panel (a) we label the different bounds that rule out the
      parameter space. This colour coding holds true for all the
      plots. In panel (b)-(d) the WMAP allowed regions are plotted in
      varying colours. The legend in panel (b) links the colour to the
      degree of fine-tuning. EW fine-tuning is represented by contours
      in panels (c) and (d). $\bsg$ agrees with measurement at $\sigI$
      across the open parameter space but $(g-2)_\mu$ agrees with the
      Standard Model value. The SUSY spectra corresponding to these
      panels are discussed in \cite{hep-ph/0403255}.}\label{T2figs}}
\end{figure}

In Figs.\ref{T2figs}(a)-(d) we gradually switch on twisted moduli
contributions by slowly decreasing $\phi$ from $\pi/2$ while keeping
$\theta=0$. This immediately gives a non-zero mass to the 3rd family
squarks and sleptons. Writing $\phi=\pi/2-\delta$, for small $\delta$
we can write the 3rd family scalar mass:

\begin{equation}
m_{0,3}\approx\frac{\delta}{\sqrt{2}}m_{3/2}\label{m03}
\end{equation}

In Fig.\ref{T2figs}(a) the parameter space of $\tanb<10$ is entirely
closed off by LEP bounds on the stau or the stau being the LSP. As we
reduce $\phi$, we give a soft mass to the stau and so increase its
physical mass, helping to satisfy the LEP bound and push its mass
above that of the $\neut$. In Figs.\ref{T2figs}(c),(d) the stau LEP
bound is no longer important. The remaining LEP bounds are the Higgs
for low $\tan\beta$ and the lightest neutralino for $\mgr<270\text
{ GeV}$.  Large $\tan\beta$ is ruled out by a failure of REWSB
($\mu^2<0$) and the stau being the LSP.

There are 4 distinct regions that satisfy dark matter bounds in the
T-moduli dominated limit. Alongside the region in which the stau is
the LSP, there is a corresponding dark matter strip in which the stau
is close in mass to the neutralino and $\neut-\stau$ co-annihilation
reduces $\DM$ to the observed value.  This is visible in
Figs.\ref{T2figs}(b)-(d) at $\mgr>450\text{ GeV}$.  For lower values
of $\mgr$, the stau is light enough that $\neut\neut\rightarrow
\tau^+\tau^-$ via t-channel stau exchange is enhanced to the point
that it alone can account for the observed dark matter density. This
is the stau analogue of the bulk region found in
\cite{hep-ph/0603095}. As we reduce $\mgr$, we are also reducing the
mass of the LSP. Before the LEP bounds close off the parameter space
there are regions in which $2m_{\neut}=m_{Z,h}$. These lie at
$\mgr=310\text{ GeV}$ and $\mgr=400\text{ GeV}$ respectively. In these
regions, the lightest neutralino can annihilate via a real on-shell
$Z$ or $h^0$.

Each of these regions has a distinct measure of fine-tuning. The
biggest surprise is the stau co-annihilation strip, shown in grey. In
contrast to the stau co-annihilation strips studied in
\cite{hep-ph/0603095}, this co-annihilation strip exhibits fine-tuning
$\DeltaO>100$. This is an order of magnitude increase over previous
stau co-annihilation regions.  The reason for this is the extreme
sensitivity to $\phi$ highlighted by Eq.\ref{m03}. In previous studies
the soft stau mass was so light that loop corrections from the
gauginos dominated the determination of its low energy mass. This
reduced the sensitivity to variations in the soft stau mass and
resulted in the low energy stau and neutralino masses being
correlated. In this model, the extreme sensitivity of the stau soft
mass to $\phi$ (for $\phi=1.47$, a $10\%$ variation in $\phi$ results
in a $150\%$ change in $m_{0,3}$) breaks this correspondence. As a
result, for $\theta=0$, the model does not have a region in which
$m_{\stau}$ and $m_{\neut}$ are correlated.

We can see this by considering the effect of changing from varying the
soft mass directly to varying it via $\phi$. Under a change of
variables:
\begin{equation}
  \DeltaO_{\phi}=\sum_{a_{MSSM}}\frac{\phi}{a_{MSSM}}\frac{\partial
    a_{MSSM}}{\partial\phi} \DeltaO_{a_{MSSM}} \label{varChange}
\end{equation}
When $\theta=0$, the coefficient of proportionality between
$\DeltaO_{\phi}$ and $\DeltaO_{m_{0,3}}$ is $\phi\tan\phi$, so as
$\phi\rightarrow\pi/2$, $\DeltaO\rightarrow \infty$. This dramatically
demonstrates the model dependence of fine-tuning.

Eq.\ref{varChange} is exact and a similar change of variables
can be performed to find all of the $\DeltaO_{a_{string}}$ in terms of
$\DeltaO_{a_{MSSM}}$. In general these expressions are large and not
particularly informative. However in cases such as that of the $\stau$
coannihilation region, we can use Eq.\ref{varChange} to understand the
change in the fine-tuning. 

The bulk region is shown in red in Figs.\ref{T2figs}(c),(d)
corresponding to $\DeltaO$ of order 10. This tuning is entirely from
$\phi$. In \cite{hep-ph/0603095}, the tuning of the bulk region came
equally from $\DeltaO_{M_1}$ and $\DeltaO_{m_0}$ where $m_0$ was the
soft mass of the slepton that mediated t-channel annihilation. In
\cite{hep-ph/0603095} the total tuning of the bulk region was found to
be low, $\DeltaO\approx 1$. When we change variables from $a_{MSSM}$ to
$a_{string}$, for $\delta\approx 0.1$, $\theta=0$, Eq.\ref{varChange}
gives $\DeltaO_{\phi}\approx 10 \DeltaO_{m_{0,3}}$ in the bulk region.
This explains the order of magnitude increase in the tuning.

Finally we consider the resonances. The lower edge of the Higgs
resonance exhibits a tuning $\DeltaO\approx 50$ whereas the edge at
larger $\mgr$ is so steep that the scan has failed to resolve it. What
we can see of it exhibits tuning well in excess of 100. In contrast
the $Z$ resonance exhibits relatively low fine-tuning. This is because
annihilation via an s-channel $Z$ is inefficient and provides only a
small contribution to the total annihilation cross-section. This is
because the $Z$ is spin 1, whereas the neutralino is a spin $1/2$
Majorana fermion. This means that in the $v_{\neut}\rightarrow 0$
limit, the annihilation cross-section via on-shell $Z$ production
becomes negligible. As this contribution is small, it hardly affects
the dark matter fine-tuning.

The electroweak fine-tuning is shown by contours on the open parameter
space. As we noted in the previous section, electroweak fine-tuning
depends closely on the largest 3rd family masses. As we can access low
$\mgr$ for large $\phi$, we end up with electroweak fine-tuning
$\mathcal{O}(100)$, similar to the lowest electroweak fine-tuning
found in the MSSM.

\subsection{Switching on the dilaton.}

\begin{table}
  \begin{center}
    \begin{tabular}{|l|r|}
      \hline
      Soft Mass & Value \\
      \hline
      $m_0^2$ & $-0.5~\mgr^2$\\
      $m_{0,3}^2$ & $\mgr^2$\\
      $m_{H}^2$ & $-2~\mgr^2$\\
      $M_i$ & $0$ \\
      $A$ & $0$\\
      \hline
    \end{tabular}
  \end{center}
  \caption{\footnotesize{The soft masses in the dilaton ($S$) dominated
    limit, $\theta=\pi/2$. This limit is characterised by
    vanishing gaugino masses and negative Higgs 
    $(\text{mass})^2$.}\label{Stab}}
\end{table}

In the limit of dilaton dominated SUSY breaking, $\theta=\pi/2$ the
soft mass terms take the form shown in Table \ref{Stab}.  This
structure of soft masses gives rise to a plethora of problems.
Firstly, negative soft sfermion mass squareds will result in tachyons.
Secondly massless gauginos are ruled out by LEP. However the biggest
problem lies in the Higgs sector. If the soft term $m_H^2$ is
negative we run the risk of breaking electroweak symmetry at the GUT
scale. This happens when $m_H^2+\mu^2<0$ at the GUT scale. We steer
clear of such regions by constraining our parameters to give
$m_H^2>0$. This allows us to impose the limit $0<\theta<0.6$.

When we consider the maximum allowed dilaton contribution, there are
two interesting limits. For $(\theta=0.6,\phi=0)$ we have $(S,Y_2)$ SUSY
breaking. When $(\theta=0.6,\phi=\pi/2)$ we have $(S,T_i)$ SUSY breaking.

For $\phi=0$, dark matter is still Wino and so cannot reproduce the
observed dark matter density. The only change is that we can access
large values of $\tan\beta$. Therefore we cannot have a model in which
there is no T-moduli contribution to SUSY breaking and reproduce the
observed dark matter density.

\begin{figure}
  \begin{minipage}{0.49\textwidth}
    \scalebox{0.7}{\includegraphics{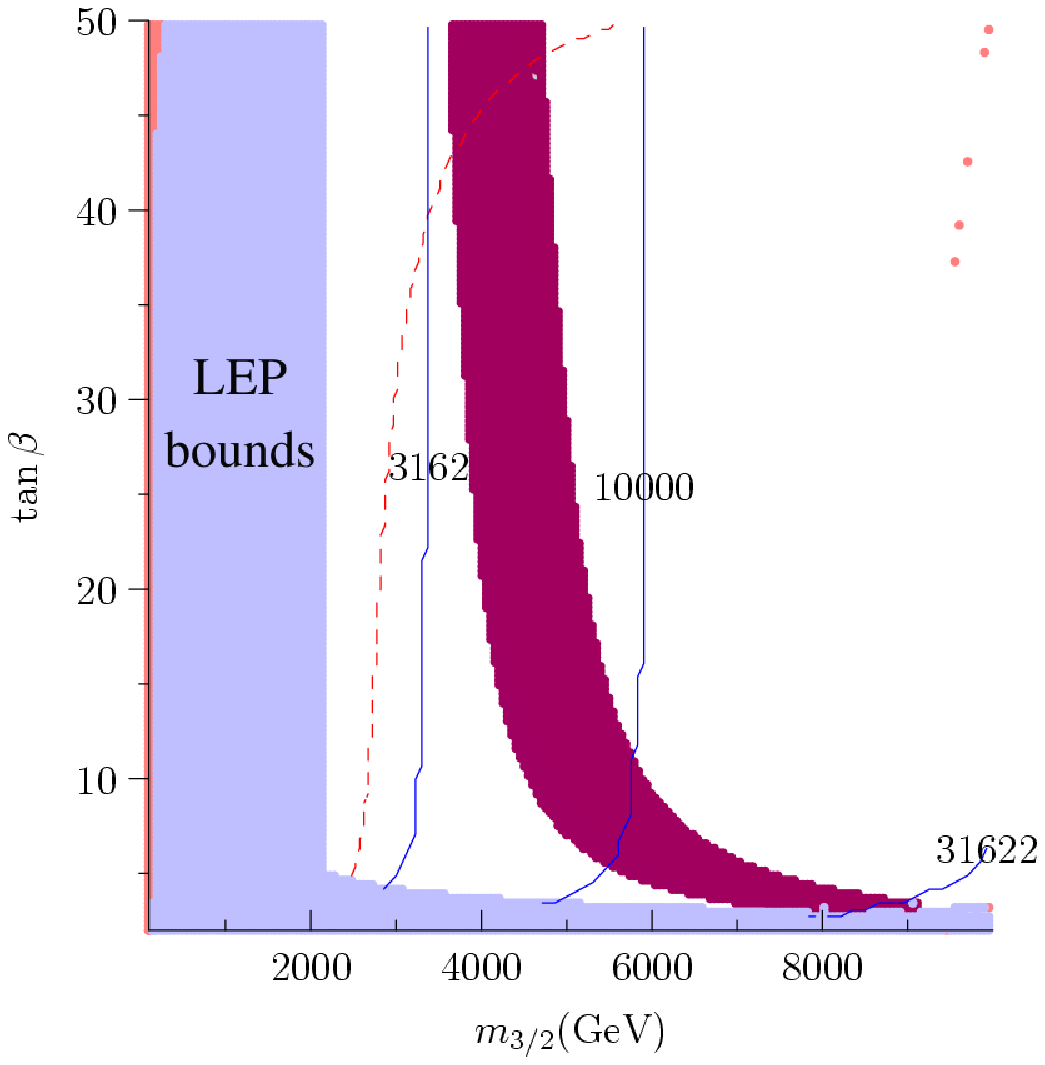}}
  \end{minipage}
  \hspace*{0.5cm}
  \hfill
  \begin{minipage}{0.49\textwidth}
    \scalebox{0.7}{\includegraphics{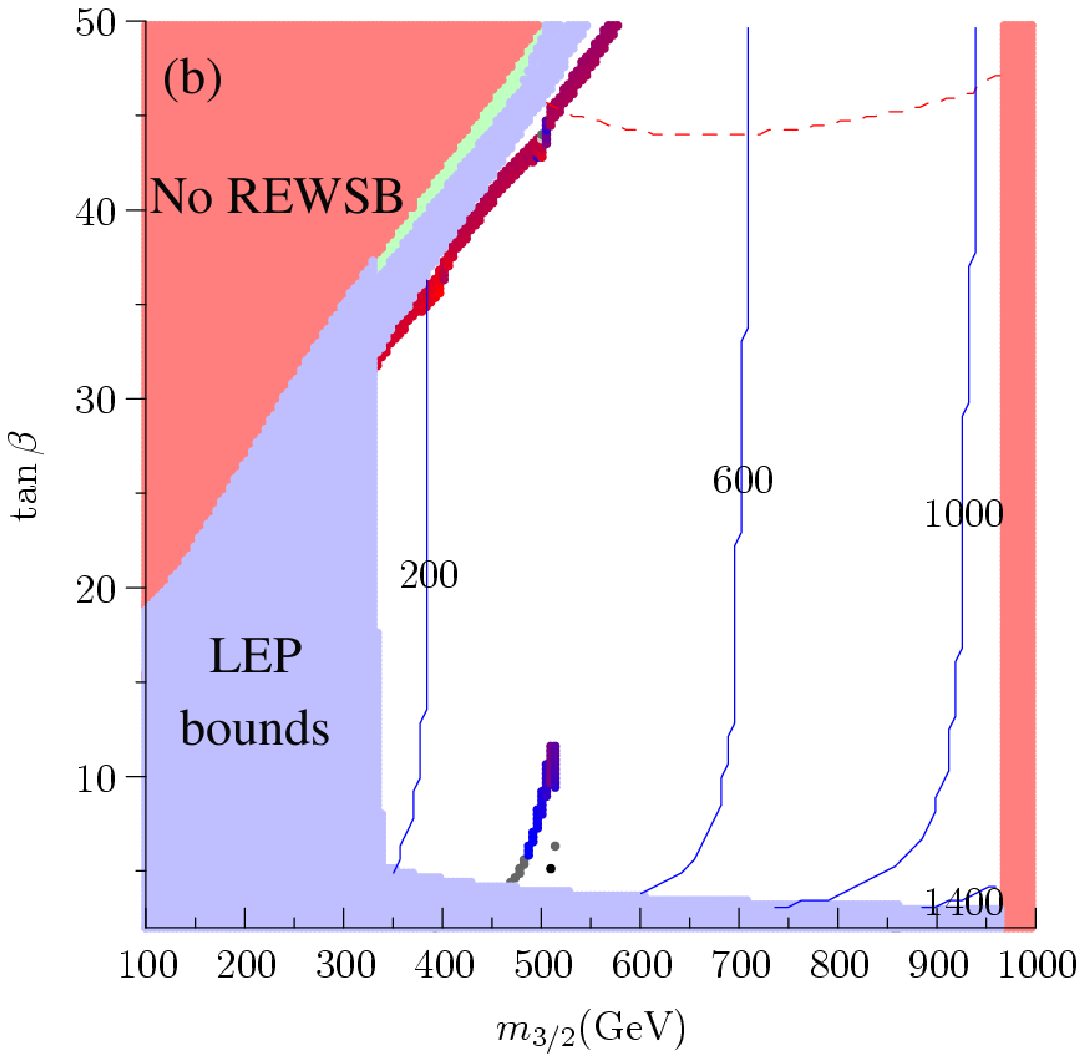}}
  \end{minipage}
  \caption{\footnotesize{Here we show the maximum dilaton contribution $\theta=0.6$.
      For larger values of $\theta$, $m_{H_{1,2}}^2<0$ at the GUT
      scale. The regions that satisfy dark matter constraints are
      plotted in varying colours to represent the required quantity of
      fine-tuning. This colour coding is as per the legend in
      Fig.\ref{T2figs}(b). The electroweak fine-tuning is represented
      by contours in the open parameter space. The $\bsg$ $\sigI$
      limit is plotted as a red dashed line. In panel (a) $\phi=0.06$,
      here we have maximally tempered Bino/Wino/Higgsino dark matter,
      plotted in purple. In panel (b) $\phi=\pi/2$, the limit in which
      there is no twisted moduli ($Y_2$) contribution. Again
      $(g-2)_\mu$ agrees with the Standard Model.}\label{Sfigs}}
\end{figure}

In Fig.\ref{Sfigs}(b) $\phi=\pi/2,~\theta=0.6$ giving $M_1<M_2$ and
hence the LSP is Bino. By introducing non-zero $\theta$ we increases the
stau mass and avoid the LEP bounds on the stau that ruled out
$\theta=0$, $\phi=\pi/2$.  It is only for large $\tan\beta$ that the
stau is light enough to contribute to neutralino annihilation via
t-channel $\stau$ exchange. As before this region is shown in red,
corresponding to $\DeltaO\approx 10$. As we can still access low
$\mgr$, there exists a region in which the neutralinos can annihilate
via the production of a real on-shell $h^0$ or $Z$. The $Z$ resonance
shows up as a small blip in the bulk region at $\mgr=400\text{ GeV}$.
The $h^0$ resonance appears as a highly tuned region (dark blue) in
the stau bulk region around $\mgr=500\text{ GeV}$ and also at
$\tanb=5-10$. For $\tanb=10-40$, even resonant annihilation via
on-shell Higgs production is not enough to suppress the dark matter
density.

As we steadily decrease $\phi$, the staus increase in mass removing
the stau bulk region. Small $\phi$ also reduces the gaugino masses,
requiring ever larger values of $\mgr$ to satisfy LEP bounds. There is
no change in the dark matter phenomenology until $\phi=0.06$, when the
neutralino acquires a large Wino component. In Fig.\ref{Sfigs}(a) we
display this region of parameter space. Here $M_1\approx M_2\approx
\mu$ at the low energy scale, resulting in maximally tempered
Bino/Wino/Higgsino dark matter as proposed in \cite{hep-ph/0603095}.
This in turn gives a wide dark matter annihilation strip shown in
purple that corresponds to $\DeltaO=23$. This tuning arises from the
soft mass sensitivity to $\phi$. This dependence is understandable as
it is $\phi$ that determines the size of the Bino and Wino
contributions to the lightest neutralino.

The electroweak fine-tuning is dependent upon the size of $\mgr$.
Therefore Fig.\ref{Sfigs}(b) exhibits low $\DeltaEW$ in agreement with
Fig.\ref{T2figs} and Fig.\ref{Sfigs}(a) exhibits large $\DeltaEW$ as
in Fig.\ref{Y2figs}.

\section{Conclusions}
\label{Conc}
We have used the measured dark matter relic density to constrain a
semi-realistic type I string model. In the model considered
supersymmetry breaking arises from F-terms of moduli fields
parameterised in terms of Goldstino angles, which automatically gives
rise to non-universal soft third sfamily and gaugino masses, which
precisely corresponds to the type of non-universality assumed in the
MSSM \cite{hep-ph/0603095}. We have studied 
fine-tuning in the string model for both
electroweak symmetry breaking and dark matter. We have
found that dark matter constraints close off much of the parameter
space of the type I string model, for example the benchmark points
suggested in \cite{hep-ph/0403255} are either ruled out ($\DM\gg
\Omega^{WMAP}_{CDM}h^2$) or disfavoured ($\DM\ll
\Omega^{WMAP}_{CDM}h^2$). However, by performing a comprehensive
scan over the parameter space, we found successful regions of dark
matter within the string model. Some of these mirror regions found in
the non-universal MSSM studies in \cite{hep-ph/0603095}. When we
consider fine-tuning, some regions exhibit degrees of fine-tuning in
agreement with the previous results while others vary significantly.
The results are summarised in Table \ref{SumTab}.

\begin{table}
  \begin{center}
    \begin{tabular}{|l|l|l|l|l|l|}
      \hline
      Dark Matter Region & $\theta$ & $\phi$ & $\mgr$(TeV) & Typical $\DeltaO$ & Typical $\DeltaEW$\\
      \hline
      Higgsino/Bino & 0-0.6 & $<0.4$ & 1-10 & 60 & $>3000$\\
      Wino/Bino & 0-0.6 & $\approx 0.06$ & 1-3 & 10 & $300-3000$\\
      Bino/Wino/Higgsino & 0-0.6 & $\approx 0.06$ & 2-5 & 10-20 & 1000-6000\\
      $\tilde{\tau}$-co-annihilation & 0-0.6 & $>0.8$ & 0.4-0.9 & 100 & 500-800\\
      t-channel $\tilde{\tau}$ exchange & 0-0.6 & $>0.8$ & 0.25-0.45 & 10 & 100-200\\
      $h^0$ resonance & 0-0.6 & $>0.4$ & $\approx 0.4$ & $>80$ & 200\\
      $Z^0$ resonance & $<0.3$ & $>0.4$ & $\approx 0.3$ & 4-20 & 130\\      \hline
    \end{tabular}
  \end{center}
  \caption{\footnotesize{A summary of the successful regions of
 parameter space in the type I string model considered here that satisfy
      experimental bounds on the dark matter density with
      corresponding typical values of $\DeltaO$ and $\DeltaEW$.}\label{SumTab}}
\end{table}

From Table \ref{SumTab} it can be seen that the observed dark matter
density tightly constrains the available parameter space. For
$\phi>0.07$, without unusual contributions to the annihilation
cross-section the model predicts an overabundance of dark matter that
would over close the universe. Equally for $\phi<0.05$, the LSP is
Wino and the model predicts a dark matter abundance orders of
magnitude less than that observed. By imposing dark matter constraints
we have ruled out the benchmark points proposed in
\cite{hep-ph/0603095}. Instead, we propose a benchmark point within
the region of lowest fine-tuning, the stau bulk region combined with
on-shell $Z$ production. The SUSY spectrum of this point is presented
in Table \ref{Bench2}.

\begin{table}[e]
\begin{center}
\begin{tabular}{|c||c|} 
  \hline
   Point & $A'$  \\ \hline
   $\theta$ & 0 \\
   $\phi$ & $3\pi/8$ \\
   $m_{3/2}$ & 310 \\ 
   $\tan\beta$ & 13 \\ \hline\hline
   $m_{h^0}$     & 115 \\
   $m_{A^0}$     & 550 \\
   $m_{H^0}$     & 550 \\ 
   $m_{H^{\pm}}$ & 556 \\ \hline
   $m_{\tilde{\chi}_1^0}$     & 44.5 \\
   $m_{\tilde{\chi}_2^0}$     & 213 \\
   $m_{\tilde{\chi}_1^{\pm}}$ & 213 \\
   $m_{\tilde{g}}$            & 930 \\ \hline
   $m_{\tilde{t}_1}$ & 546 \\
   $m_{\tilde{t}_2}$ & 757 \\
   $m_{\tilde{c}_L}$, $m_{\tilde{u}_L}$ & 3390 \\
   $m_{\tilde{c}_R}$, $m_{\tilde{u}_R}$ & 3390 \\
    \hline
   $m_{\tilde{b}_1}$ & 687 \\
   $m_{\tilde{b}_2}$ & 739 \\
   $m_{\tilde{s}_L}$, $m_{\tilde{d}_L}$ & 3390 \\
   $m_{\tilde{s}_R}$, $m_{\tilde{d}_R}$ & 3390 \\ \hline
   $m_{\tilde{\tau}_1}$ & 104 \\
   $m_{\tilde{\tau}_2}$ & 222 \\
   $m_{\tilde{\mu}_L}$, $m_{\tilde{e}_L}$  & 3290 \\
   $m_{\tilde{\mu}_2}$, $m_{\tilde{e}_2}$  & 3280 \\ 
    \hline 
   $m_{\tilde{\nu}_e}$, $m_{\tilde{\nu}_{\mu}}$ & 3290 \\
   $m_{\tilde{\nu}_{\tau}}$ & 197 \\ \hline
   LSP & $\tilde{\chi}_1^0$  \\ 
    \hline     
\end{tabular}
\end{center}
\caption{\footnotesize{Sample spectra for benchmark point $A'$
    corresponding to a point in Fig.~\ref{T2figs}(d) at
    $m_{3/2}=310$ GeV and $\tan \beta = 13$. At this point we
    satisfy WMAP bounds on the dark matter density, $\bsg$ and all
    present mass bounds. This point requires a tuning to achieve
    electroweak symmetry breaking: $\DeltaEW=125$, and a tuning to
    agree with WMAP: $\DeltaO=3.9$. The annihilation of neutralinos in
    the early universe is due to $40\%~\neut\neut\rightarrow
    \tau^+\tau^-$ via t-channel $\stau$ exchange and
    $60\%~\neut\neut\rightarrow f \overline{f}$ via the production of
    an on-shell $Z$. All masses are in GeV.} \label{Bench2}}
\end{table}

In addition to constraining our models, we have been able to study how
fine tuning varies between the MSSM studied in \cite{hep-ph/0603095}
and a type I string model of SUSY breaking, which was one of our main
motivations for this study.  From Table \ref{SumTab} it can be seen
that, in the string model, the lowest dark matter fine-tuning exists
in the bulk region, corresponding to t-channel $\stau$ exchange.  The
$Z$ resonance, the well tempered Bino/Wino and the maximally tempered
Bino/Wino/Higgsino regions also have low dark matter fine-tuning.  Of
these, the lowest electroweak fine-tuning arises in the bulk
(t-channel $\stau$ exchange) and $Z$ resonance regions.  These results
are consistent with the conclusions based on the previous MSSM
analysis, although the bulk region in the MSSM corresponding to first
and second family slepton exchange cannot be accessed in the string
model as discussed.  Thus in most cases the degree of fine-tuning is
found to be the same order of magnitude as found for similar dark
matter regions within the MSSM. However this is not always the case.
Whereas the well tempered Higgsino/Bino region in Table \ref{SumTab}
continues to be highly fine-tuned as in the MSSM, the well tempered
Bino/Wino in Table \ref{SumTab} has a fine tuning of about 10 as
compared to the MSSM value of about 30, making this scenario more
natural in the framework of string theories such as the one considered
here.

In some cases there is a sharp disagreement between the fine tuning
calculated in the MSSM and in the string model, for example in the
stau co-annihilation region. Due to the form of the SUSY breaking in
this model, the stau mass, and so the dark matter density, is very
sensitive to $\phi$ which leads to an order of magnitude increase in
the dark matter fine-tuning in the string model as compared to the
MSSM, making this region less natural in the string model. This can be
understood via Eq.\ref{varChange} which shows that, through a general
change of variables, the variation of the fine-tuning between a
general MSSM model and a string model can be calculated.  In principle
a similar change of variables is responsible for the all the
differences in fine tuning calculated in the MSSM and the string
model. In practice however, such a change of variables is not
analytically tractable, and numerical methods such as those used in
the present paper are required in order to obtain quantitative
results. However the results in this paper indicate a general strategy
for reducing fine tuning within string models, namely to search for
string models that minimise the coefficients of the tuning measures.
This in turn will minimise $\DeltaO$, providing more natural dark
matter than the MSSM for a given region of parameter space.  Such a
strategy could also be employed to reduce electroweak fine tuning once
the solution to the $\mu$ problem is properly understood within the
framework of string theory.

\section{Acknowledgements}

The authors gratefully acknowledge the hospitality of the CERN theory
division. JPR acknowledges a PPARC Studentship.

\end{document}